\documentclass{article}
\usepackage{spconf,amsmath,graphicx,comment,amsfonts, paralist,algorithm,algorithmic,amsthm,amssymb}
\usepackage{tikz}
\usepackage{subcaption}
\usepackage[european resistor, european voltage, european current]{circuitikz}
\usetikzlibrary{arrows,shapes,positioning,snakes,calc}
\usetikzlibrary{decorations.markings,decorations.pathmorphing,decorations.pathreplacing}
\usetikzlibrary{calc,patterns,shapes.geometric,topaths,fit}

\newcommand{\blue}[1]{\textcolor{blue}{#1}}

\newcommand{\probe}{\alpha}
\newcommand{\response}{\beta}
\newcommand{\utility}{u}

\newcommand{\reals}{\mathbb{R}}
\newcommand{\dataset}{\mathcal{D}}

\renewcommand{\time}{k}

\newcommand{\horizon}{K}

\newcommand{\probedim}{m}

\newcommand{\argmax}{\operatorname{argmax}}

\newtheorem{theorem}{Theorem}

\newcommand{\Prob}{\mathbb{P}}

\newcommand{\radarUtility}{\utility_R}

\newcommand{\jammerUtility}{\utility_J}
\newcommand{\jammerUtilityest}{{\hat{\utility}_J}}
\newcommand{\jammerUtilityestiter}{{\widehat{\utility}_{J,\time}}}

\newcommand{\Variance}{\Sigma}

\newcommand{\timeIndexOne}{n}
\newcommand{\timeIndexTwo}{k}

\newcommand{\updateparam}{\operatorname{\texttt{IRL-Update}}}
\newcommand{\param}{\theta}
\newcommand{\paramset}{\Theta}

\newcommand{\AFT}{\operatorname{AFT}}
\newcommand{\margin}{\mathcal{M}}
\newcommand{\paramestiter}{\hat{\param}_\time}

\newtheorem{protocol}{Protocol}
\usepackage{quoting}
\quotingsetup{vskip=1pt}

\tikzstyle{block} = [draw, rectangle,
    minimum height=3em, minimum width=6em]
\tikzstyle{blockd}=[draw, blue, dashed, rectangle, minimum height=5em, minimum width=19em]
\tikzstyle{blockj}=[draw, red, dashed, rectangle, minimum height=5em, minimum width=19em]
\title{Adaptive ECCM for Mitigating Smart Jammers}
\name{ Kunal Pattanayak$^\ast$,Shashwat Jain, \sthanks{V. Krishnamurthy, K. Pattanayak and S. Jain are with the School of Electrical and Computer Engineering, Cornell University, Ithaca,	NY, 14853 USA. e-mail: vikramk@cornell.edu, kp487@cornell.edu, sj474@cornell.edu. $^\dag$ C. Berry is with Lockheed Martin Advanced Technology Laboratories, Cherry Hill, NJ, 08002 USA. e-mail: christopher.m.berry@lmco.com. This research was supported in part by a research contract from  Lockheed Martin and  the Army Research Office grant W911NF-21-1-0093.}, Vikram Krishnamurthy$^\ast$ and Christopher Berry$^\dagger$}
\address{$^\ast$ Electrical and Computer Engineering, Cornell University, USA\\
$^\dag$ Lockheed Martin Advanced Technology Laboratories, USA. }

\begin{document}
%\ninept
%
\maketitle
\begin{abstract}
{This paper considers adaptive radar electronic counter-counter measures (ECCM) to mitigate ECM by an adversarial jammer. Our ECCM approach models the jammer-radar interaction as a Principal Agent Problem (PAP), a popular economics framework for interaction between two entities with an information imbalance. In our setup, the radar does not know the jammer's utility. Instead, the radar learns the jammer's utility adaptively over time using inverse reinforcement learning. The radar's adaptive ECCM objective is two-fold (1) maximize its utility by solving the PAP, and (2) estimate the jammer's utility by observing its response. Our adaptive ECCM scheme uses deep ideas from revealed preference in micro-economics and principal agent problem in contract theory. Our numerical results show that, over time, our adaptive ECCM both identifies and mitigates the jammer's utility.
} 
\end{abstract}
\section{Introduction}
\label{sec:intro}
%Demand Information Asymmetry and Cost Information Asymmetry.
{\em Electronic Countermeasures} (ECM) are widely used by jammers to degrade radar's measurement accuracy in a shared spectrum environment. To mitigate the impact of ECM, modern radars implement {\em Electronic Counter-Countermeasures} (ECCM). We consider a target-tracking cognitive radar that maximizes its signal-to-noise ratio (SNR) and is also aware of an adversarial jammer trying to mitigate the radar's performance. At the start of the radar-jammer interaction, the radar has zero information about the jammer's strategy. The radar learns the jammer's ECM strategy over time via repeated radar-jammer interactions while ensuring its measurement accuracy lies over a specified threshold - we term this as {\em adaptive ECCM}. A list of standard ECM and ECCM strategies are summarized in \cite{John78} and \cite{Skol08}.

This paper formulates the radar's ECCM objective as a {\em Principle Agent Problem} (PAP), wherein the radar gradually learns the jammer's objective using {\em Inverse Reinforcement Learning} (IRL). We assume the radar possesses IRL capability and can learn the jammer's utility, while the jammer is a naive agent - it only maximizes its utility.
Reconstructing agent preferences from a finite time series dataset is the central theme of revealed preference in micro-economics~\cite{Afr67}, \cite{Var12}. In the radar context, the radar uses the celebrated result of Afriat's theorem~\cite{Afr67} to estimate the jammer's utility over time. This imbalance of information is termed as information asymmetry in literature widely studied in micro-economics \cite{dierkens1991information}\cite{tsvetkov2014information}. The Principal Agent Problem (PAP) is a well-known framework that {\em mitigates} information asymmetry. The PAP \cite{1995:AM} has been studied extensively in micro-economics for appropriate contract formulation between two entities in  labor contracts \cite{2005:GM}, insurance market \cite{2003:MH}\cite{rauchhaus2009principal}, and differential privacy \cite{2014:CD-AR} in machine learning. 

Our choice of PAP is motivated by its flexibility to accommodate additional constraints on the information of the radar and the jammer. Such capabilities have been capitalized upon in prior ECCM literature \cite{AG2022}. However, our work generalizes \cite{AG2022} in that the radar now has to learn the jammer's utility over time using IRL. Similar approaches for mitigating information asymmetry exist in literature, for example, in consumer economics \cite{KA2015} where the seller maximizes its profit by solving a leader-follower problem. In \cite{APR21-STGame}, the seller maximizes his utility by modelling his interaction with the consumer as a Stackelberg game.

%We model the radar as the principal and the adversarial jammer as the agent. The information asymmetry aspect is due to principal's uncertainty of agent's utility. The principal only observes the agent's responses and estimates the utility of the agent using IRL. It then writes a contract with suitable incentives to induce response from the agent that would maximize its utility.

%In the PAP, the radar ensures an optimal balance between measurement accuracy and \textit{measurement cost}: cost to generate radiation power for the pulse to probe the target. It is important that we incorporate measurement cost while designing ECCM as a large number of measurements have to be made for continuous monitoring of targets. Similarly, the jammer has to consider a \textit{jamming cost} for generating radiation noise power. Again it is important to include it in the jammer's utility function as the jamming has to be done continuously to reduce the radar's tracking accuracy. We study through numerical examples a radar's ECCM problem when the radar and the jammer have information asymmetry.

\section{Radar-Jammer Interaction for Adaptive ECCM}
\label{sec:radar_jammer_ew}
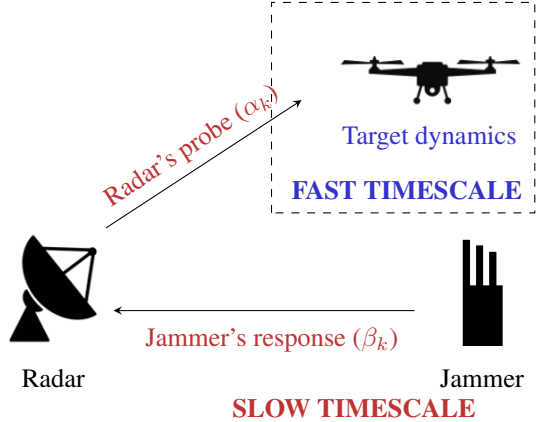
\begin{figure}
	\begin{center}
		\begin{tikzpicture}
			\tikzstyle{arrow} = [->,>=stealth];
			\definecolor{intermediate}{rgb}{0.75,0.2,0.2};
			\definecolor{fast}{rgb}{0.2,0.2,0.8};
			\definecolor{slow}{rgb}{0.75,0.2,0.2};
			%\draw [dashed] (0,-0.2) rectangle(8,8);
			%\draw [dashed] (0.1,0.2) rectangle(7.8,7);
			\draw [dashed] (4.1,4) rectangle(7.6,6.8);
			\node[inner sep=0] (target) at (6.2,5.8)
			{\includegraphics[width=0.15\textwidth]{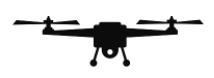}};
			\node[draw, align=right, draw=none, color=fast] at (6.2,5) {Target dynamics};
			\node[inner sep=0] (radar) at (1.2,3)
			{\includegraphics[width=0.1\textwidth]{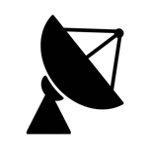}};
			\node[draw, align=right, draw=none] at (1.2,1.8) {Radar};
			\node[inner sep=0] (jammer) at (6.9,2.94)
			{\includegraphics[width=0.03\textwidth]{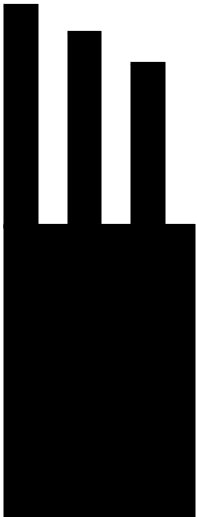}};
			\node[draw, align=right, draw=none] at (6.9,1.8) {Jammer};
			\draw [arrow] (1.9,3.8) -- node[anchor=south, sloped, color=intermediate] {Radar's probe ($\probe_{\timeIndexTwo}$)} (4.5,5.5);
			%\draw [arrow] (4.7,5.2) -- node[anchor=north, sloped, color=intermediate] {Echo} (2.1,3.5);
			\draw [arrow] (6,2.7) -- node[anchor=north, sloped, color=intermediate, align=right, xshift=3,yshift = -3] {{Jammer's response ($\response_{\timeIndexTwo}$)}} (2,2.7);
			%\draw [arrow] (3.1,2.7) -- node[anchor=north, sloped, color=intermediate, xshift=12] {\parbox{2cm}{Radar's\\observation $\radarObservation$}} (2,2.7);
			%\node[draw, align=center, thick] at (4,2.7) {Ambient\\noise};
			\node[draw, align=right, draw=none, color=fast, font=\bfseries] at (5.9,4.3) {FAST TIMESCALE};
			\node[draw, align=right, draw=none, color=intermediate, font=\bfseries] at (5.2,1.4) {SLOW TIMESCALE};
			%\node[draw, align=right, draw=none, color=slow, font=\bfseries] at (6,0.1) {SLOW TIMESCALE};
			%\node[draw, align=right, draw=black, very thick, text=slow] at (6.3,7.5) {Target maneuver};
			%\draw [-stealth, very thick] (6.25,7.24) -- (6.25,6.5);
		\end{tikzpicture}
	\end{center}
	\caption{Two timescale radar-jammer problem for tracking a target. Target dynamics/transient occurs on the fast timescale; radar-jammer PAP is solved at the slow timescale.}
	\label{fig:block_diagram}
\end{figure}

In this section, we formulate the radar, jammer and target interaction as shown in Fig.~\ref{fig:block_diagram}.
The radar and jammer interact while the target evolves independently. The main idea is that the radar exploits this interaction to mitigate the effect of ECM by using ECCM. For simplicity we only consider a single target. %The radar tracks the target using a Bayesian tracker \cite{1998:YB}. The adversarial jammer (mounted on a dedicated ECM ship \cite{2014:NR} or  aircraft) injects jamming power into the environment as an electronic countermeasure (ECM) to decrease the measurement accuracy of the radar. In response, the radar varies its pulse power as an electronic counter-countermeasure (ECCM) to mitigate the presence of the adversarial jammer.
%\footnote{In our setup, the jammer, and target are independent entities. \cite{2008:LX-et-al}, \cite{2018:CW-et-al}, \cite{2020:QL-et-al} also study a similar model with independent targets and jammers. In an alternative setup (not considered here), the jammer is mounted on the target \cite{2012:XS-et-al}. This results in the radar's observation $\radarObservation$ of the jamming power $\jammerStrategy$ also being dependent on the target's kinematics.}
We formulate the scenario in two timescales. Let $\timeIndexOne\in \{0,1,2,\ldots\}$ and $\timeIndexTwo\in \{0,1,2,\ldots\}$, denote the time index for fast and slow timescale, respectively. In fast timescale, the target evolves in linear, time-invariant dynamics and additive Gaussian noise. On the slow timescale, the radar and the jammer participate in an EW and update their ECCM and ECM, respectively.

%We consider the two frameworks below. Our first framework, discussed in Sec.\ref{sec:fast_timescale} considers ECCM against barrage jamming (ECM) by an adversarial jammer; the jammer and the target operate independently. Our second framework discussed in Sec.\ref{sec:example-2} considers ECCM  against deception jamming by an adversarial jammer. Here the jammer exploits information about the target's state to jam the radar (for example, the jammer is mounted on the target). Finally, for both these frameworks, Sec.\ref{sec:intermediate_timescale} formulates the PAP for the optimal ECCM and ECM strategy of the radar and jammer.

%\subsection*{ECCM against barrage jamming}
\label{sec:fast_timescale}
\subsubsection*{Target dynamics and Radar's measurement}
\label{sec:target-dynamics}
The targets evolve on the fast time scale $\timeIndexOne$.  We use the standard linear Gaussian model\,\cite{1998:YB}  for the kinematics of the target and initial condition $x_0$:
\begin{equation} 
	\label{eq:target_model}
	\begin{split}
		x_{\timeIndexOne+1}&=A\,x_{\timeIndexOne}+w_{\timeIndexOne},\quad x_{0} \sim \mathcal{N}(\hat{x}_0,\Variance_{x_0})\\ 
        y_{\timeIndexOne} & = h(x_\timeIndexOne,v_\timeIndexOne),\quad v_{\timeIndexOne}\sim\mathcal{N}\bigg(0,V\Big(\probe_\timeIndexTwo,\response_\timeIndexTwo\Big)\bigg)\\
		A &=\operatorname{diag}\left[\textbf{T},\textbf{T},\textbf{T}\right], \quad \textbf{T}=\left[\begin{array}{ll}1 & T \\ 0 & 1\end{array}\right]
	\end{split}
\end{equation}
Here, $T$ denotes the sampling time. The initial condition $\mathcal{N}(\hat{x}_0,\Variance_{x_0})$ denotes a Gaussian random vector with mean $\hat{x}_0$ and covariance $\Variance_{x_0}$. $x_\timeIndexOne\in\mathbb{R}^6$, comprised of the $x,y,z$ position and velocity, is the state of the target at time $\timeIndexOne$. The i.i.d. sequence of Gaussian random vectors $\{w_{\timeIndexOne}\sim\mathcal{N}(0,Q)\}$ models the acceleration maneuvers of the target.
Here, $h(\cdot)$ represents the radar's sensing functionalities. $V(\probe_\timeIndexTwo,\response_\timeIndexTwo)$ is the measurement noise variance of the random vector $v_k$. Here, $\timeIndexTwo$ indexes the slow timescale on which the radar and the jammer update their strategies.

%%%%%%%%%%%%%%%%%%%%%%%%%%%%%%%%%%%%%%%%%%%%%%%%%%%%%%%%%%%%%%%
%\subsubsection*{Radar's measurement model}
%\label{sec:measurement-model}
%The measurement vectors $y_{\timeIndexOne}$ of the target's state recorded at the radar are
%\begin{align}
%    \begin{aligned}
%	xx
%	\end{aligned}
%\end{align}

%\subsubsection{Bayesian Tracker and Covariance}
We make the standard assumption that the radar has a Bayesian tracker \cite{krishnamurthy2016partially}, which recursively computes the posterior distribution $\Prob(x_\timeIndexOne|y_1,\ldots y_\timeIndexOne)$ of the target state at each time $k$. The radar can also do inverse bayesian filtering for target tracking \cite{pattanayak2022can},\cite{krishnamurthy2021adversarial}.

\label{sec:bayesian-tracker}
\section{Radar-Jammer Interaction. Principal Agent Problem (PAP) with IRL}
\label{sec:PAP-IRL}
\begin{figure}[!ht]
    \centering
    \begin{tikzpicture}[auto, node distance=2cm,>=latex']
    % We start by placing the blocks
    \node[blockd](Radar){};
    \node[blockj, below of=Radar, node distance=2cm](Jammer1){};
    \node[block, below of=Radar, node distance=2cm](Jammer){Maximize $\jammerUtility$};
    \node[block,left of =Radar, node distance=2cm, align=center](IRL){IRL (Learn $\jammerUtility$\\ via Thm.\,\ref{thrm:rp})};
    \node[block, right of=Radar, node distance=2cm, align=center](Principal){PAP\\(maximize $\radarUtility$)};
    \draw[->, thick] (Jammer.west)-| node[left]{$\response_\time$} (IRL.south);
    \draw[->,thick] (Principal.south)|- node[right]{$\probe_\time$} (Jammer.east);
    \draw[->,thick] (IRL.east)--node{$\jammerUtilityestiter$}(Principal.west);
    \node[above of=Radar, node distance=1.2cm](Node1){\blue{Radar}};
    \node[below of=Jammer1,node distance=1.2cm](Node2){\textcolor{red}{Smart Jammer}};
\end{tikzpicture}
    \caption{The EECM problem, involves radar maximizing its utility based on its noisy estimate of the jammer utility. The radar estimates the utility via IRL which aids it to learn the jammer's utility adaptively. }
    \label{fig:eccm-schematic}
\end{figure}
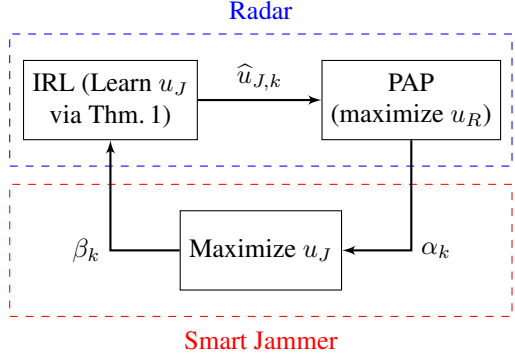

The working assumption in this paper is that the radar does not know the jammer's utility at the start of the radar-jammer interaction. The radar `learns' the jammer's utility over time via inverse reinforcement learning (IRL) as schematically illustrated in Fig.\,\ref{fig:eccm-schematic}. The radar's ECCM efficacy of mitigating the jammer improves with increasing accuracy of the radar's estimate of the jammer's utility.
% i.e.\,, the precision of the radar's estimate increases with time due to the radar's IRL block as schematically illustrated in Fig.\,\ref{fig:eccm-schematic}. We also assume the jammer chooses its jamming signal {\em after} the radar transmits its tracking signal.

\subsubsection*{Information Symmetry. Radar knows Jammer's Utility}
For simplicity, let us first consider the case where the radar completely knows the jammer's utility. This scenario is referred to as `information symmetry' in contract theory. The radar optimizes its action at time $\time$ under the assumption that the jammer is cognitive, i.e.\,, the jammer is a utility maximizer. The radar's transmitted signal at time $\time$ is given by:
\begin{protocol}[Principal Agent Problem (PAP)]\label{prtcl:PAP}
\begin{align}
\text{Radar's Probe: }&\probe_\time=\argmax~\radarUtility(\probe,\response),\label{eq:pa-general}\\
\response\in\argmax_{\bar{\response}}~\jammerUtility(\probe,\bar\response)&,~\jammerUtility(\probe,\response)  = \utility_\param(\response) +\lambda~ \probe'\response.\label{eqn:jammer_utility_rep}\\
\text{Jammer's Response: } & 
\label{eqn:abstract-cog-jammer}
    \response_\time = \argmax_{\response} \jammerUtility(\probe_\time,\response).
%&\jammerUtility(\probe,\response) = \utility(\response;\param) +\lambda~ \probe'\response.\label{eqn:jammer_utility_rep}
\end{align}
\end{protocol}
In \eqref{eq:pa-general}, $\radarUtility,~\jammerUtility$ denote the utility functions of the radar and the jammer, and $\probe_\time,~\response_\time$ denote the radar's and jammer's transmission signals at time $\time$, respectively. Observe that utilities depend on both the radar's and jammer's responses. In words, the constraint in the optimization problem \eqref{eq:pa-general} means the jammer's response for radar's signal $\probe$ maximizes the jammer's utility $\jammerUtility$, hence the feasible tuples $(\probe,\response)$ over which the radar maximizes its utility is restricted to a manifold. In \eqref{eqn:jammer_utility_rep}, $\utility_\param(\response)$ is the jammer's system utility parametrized by vector $\param$ that is independent of the radar's signal $\probe$ . In \eqref{eq:pa-general}, it is implicit that the radar knows $\param$. The radar's utility $\radarUtility$ can be a combination of tracking performance metrics, for e.g.\,, SNR, and the negative of the jammer's utility (this term mitigates the jammer). We will motivate this reasoning with a concrete example in Sec.\,\ref{sec:numerical}. Since the radar knows the jammer's utility~\eqref{eqn:jammer_utility_rep}, it can successfully ensure the jammer's utility is mitigated by a smart choice of $\probe_\time$ obtained by solving \eqref{eq:pa-general}. 

Finally, a few words on the jammer's utility structure \eqref{eqn:jammer_utility_rep}. The first term can be viewed as the jammer's system performance and is independent of the radar's action. The second term $\lambda \probe'\response$ regulates the jammer's signal power in terms of the radar's transmission signal power. Intuitively, large radar transmission power requires a large jammer transmission power for effective mitigation of the radar; see \cite{krishnamurthy2021adversarial} for the relation between $\probe'\response$ and the asymptotic covariance of the Bayesian tracker of the jammer.

\subsubsection*{Information Asymmetry. Radar learns Jammer's utility via repeated interactions}
We now consider the {\em information asymmetric} version of Protocol~\ref{prtcl:PAP}, where the radar has a noisy estimate of the radar's utility. The rest of the paper deals with the information asymmetric case. Compared to the information symmetric case, the radar's aim now is two-fold: (1) maximize its utility $\radarUtility$, and (2) estimate the jammer's utility and {\em decrease information asymmetry}. Intuitively, a more accurate estimate of the jammer's utility results in increased efficacy of radar ECCM. The radar-jammer interaction \eqref{eq:pa-general} can now be expressed as: 
% Information asymmetry for the radar will be introduced when it does not know the utility function of the jammer. In this formulation the radar only knows the approximate jammer utility $\jammerUtilityest$ and uses revealed preference to compute jammer utility. The iterative optimization problem for a given set of $(\probe_\time,\,\response_\time)_{k=1}^{\horizon-1}$ pairs is given by:
\begin{protocol}[Adaptive ECCM. Hybrid PAP with IRL] \label{prtcl:adaptive-ECCM}
Consider a radar tracking a target in the presence of an adversarial jammer. Assume the radar does not know the jammer's utility. The radar takes the following actions for all $\time>0$:\\
\underline{Step 1.} Generate signal $\probe_\time$ by solving the following PAP: 
\begin{align}
&\probe_\time =\argmax~\radarUtility(\probe,\response),~
\response\in\argmax_{\bar{\response}}~\widehat{\utility}_{J,\time-1}(\probe,\bar\response),\label{eq:pap-irl-general}\\
&\text{where }\widehat{\utility}_{J,\time-1}(\probe,\response) = \utility_{\hat{\param}_{\time-1}}(\response) +  \lambda~\probe'\response\label{eqn:pap-jammer-utility-est}.
\end{align}
In \eqref{eqn:pap-jammer-utility-est}, $\hat{\param}_{\time-1}$ is the radar's estimate of the jammer's utility parameter $\param$~\eqref{eqn:jammer_utility_rep} at time $\time$.\\
\underline{Step 2.} Receive jammer's signal $\response_\time$ and update the estimate of the jammer's utility:
\begin{align}
    \paramestiter & = \updateparam(\dataset_{\time-1}\cup\{\probe_\time,\response_\time\}),\label{eqn:param-update}\\
    \dataset_{\time-1} & = \{\probe_{\time'},\response_{\time'}\}_{\time'=1}^{\time-1}\label{eqn:dataset-def}
\end{align}
where $\updateparam(\cdot)$ is described in \eqref{eqn:IRL-update}, and $\dataset_\time$ is the radar's dataset at time $\time$. 
\end{protocol}
Protocol~\ref{prtcl:adaptive-ECCM} is our adaptive ECCM scheme illustrated in Fig.\,\ref{fig:eccm-schematic}. In Protocol~\ref{prtcl:adaptive-ECCM}, Steps 1 and 2 correspond to the PAP and IRL blocks, respectively, in Fig.\,\ref{fig:eccm-schematic}. In words, the radar first chooses its action $\probe_\time$ that maximizes its utility $\radarUtility$ subject to its current estimate of the jammer's utility. In response to radar's signal $\probe_\time$, the jammer chooses action $\response_\time$. The radar uses this new data point of $(\probe_\time,\response_\time)$ to update its estimate of the jammer's utility. The asymptotic goal of the radar is to identify that probe signal $\probe^\ast$ that satisfies:
\begin{align}
    \probe^\ast & \in \argmax_{\probe} \radarUtility(\probe,\response),\text{ such that}\nonumber\\
    \response & \in\argmax_{\response'} \jammerUtility(\probe,\response')\label{eqn:asymptotics}
\end{align}
That is, as time $\time\rightarrow\infty$, the radar knows the jammer's utility exactly and is able to effectively mitigate  ECM tactics. 
%The radar can also choose to maximize $\radarUtility - \delta ~\jammerUtility$, $\delta>0$ is the radar's spoofing parameter, in which case \eqref{eqn:asymptotics} can be viewed as a minimax problem - convergence properties of the algorithm is a subject of current research. 
%Such a setup warrants discussion from a Stackelberg game perspective and mechanism design and is the subject to future work.

\subsubsection*{IRL for Learning the Jammer's Utility}
We now describe the IRL aspect ($\updateparam(\cdot)$ module) in the adaptive ECCM outline in Protocol~\ref{prtcl:adaptive-ECCM}. Below, we present a key result from revealed preference theory, namely, Afriat's theorem~\cite{Afr67,Die12,Var12}.  Afriat's theorem is a remarkable result in micro-economics for non-parametric utility estimation from a finite dataset of probes and responses generated by a decision-maker.

In the radar context of this paper, Afriat's theorem generates a set-valued estimate of the utility parameter $\param$~\eqref{eqn:jammer_utility_rep} in the jammer's utility $\jammerUtility$. Since the radar knows {\em a priori} that the jammer's utility has a parametric form~\eqref{eqn:jammer_utility_rep}, Afriat's theorem below yields tighter conditions for utility maximization.
%As will be discussed in Theorem~\ref{thrm:online-irl}, this parametrization facilitates the radar to perform IRL using only $\BigO(\horizon)$ computations instead of $\BigO(\horizon^2)$ computations as is typical of the classical Afriat inequalities~\cite{Afr67}. 

\begin{theorem}[Afriat's theorem for Identifying Jammer Utility]\label{thrm:rp} Consider a radar tracking a target in the presence of a jammer. Suppose the parameter $\param$ in the jammer's utility~\eqref{eqn:jammer_utility_rep} belongs to a compact set $\paramset\subseteq\reals^n$, $\param$ is fixed for all $\time$ and the radar knows the value of $\lambda$ in the jammer's utility~\eqref{eqn:jammer_utility_rep}. Then, at time $\time$:\\
1. The radar checks for the existence of a feasible parameter $\param$ that satisfies \eqref{eqn:abstract-cog-jammer} by checking the feasibility of a set of inequalities:
\begin{equation}
\begin{split}
   \hspace{-0.4cm}& \text{There exists a feasible }\param'\in\paramset~\text{s.t. }\AFT(\param',\dataset_\time)\leq \mathbf{0},\\
  \hspace{-0.4cm}\Leftrightarrow  &\exists~\jammerUtility~\text{s.t. } \response_{\time'} \in \argmax_\response \jammerUtility(\probe_{\time'},\response),~\forall\time'=1,\ldots,\time,
\end{split}\label{eqn:abstract_IRL_utility}
\end{equation}
where dataset $\dataset_\time$ is defined in \eqref{eqn:dataset-def} and the set of inequalities $\AFT(\param',\dataset_\time)\leq\mathbf{0}$ is given by:
\begin{equation}\label{eqn:AFT-ineq}
\hspace{-0.11cm}    \utility_{\param'}(\response_s)-\utility_{\param'}(\response_t)+\lambda \probe_t' (\response_s-\response_t) \leq 0 \forall t,s\in\{1,\dots,\time\}.
\end{equation}
%$\dataset_\probe =\{\probe_\time'(\response_s-\response_\time)\}_{s,\time=1}^{\horizon}$.
\noindent (b) If $\AFT(\cdot,\dataset_\time)\leq\mathbf{0}$ has a feasible solution, the set-valued estimate of the jammer's utility parameter $\param$ is given by:
\begin{equation}\label{eqn:set-of-feasible-param}
\paramset_\time = \{\param':\AFT(\param',\dataset_\time)\leq\mathbf{0}\}
\end{equation}
\end{theorem}

\begin{figure}[ht]
\centering
\begin{subfigure}[b]{0.35\textwidth}
   \includegraphics[width=1\linewidth]{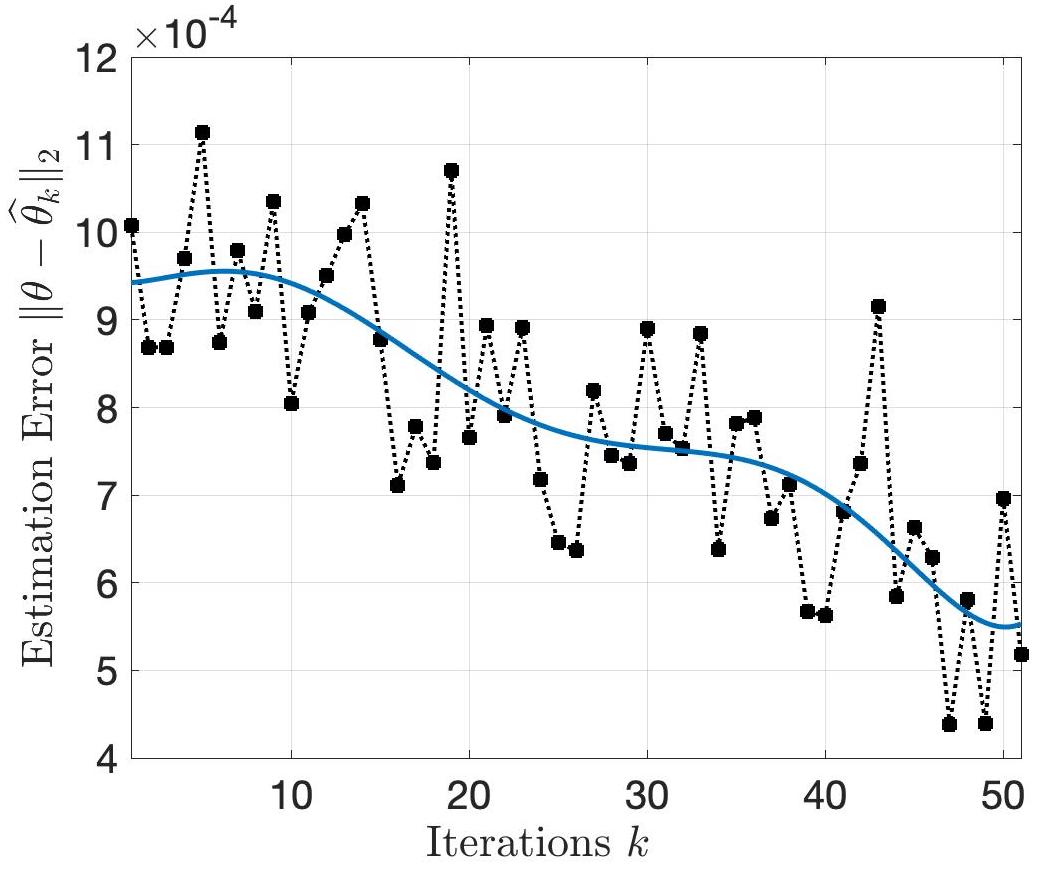}
   \caption{}
   \label{fig:est-error}
\end{subfigure}\vspace{-0.1cm}
\begin{subfigure}[b]{0.37\textwidth}
   \includegraphics[width=1\linewidth]{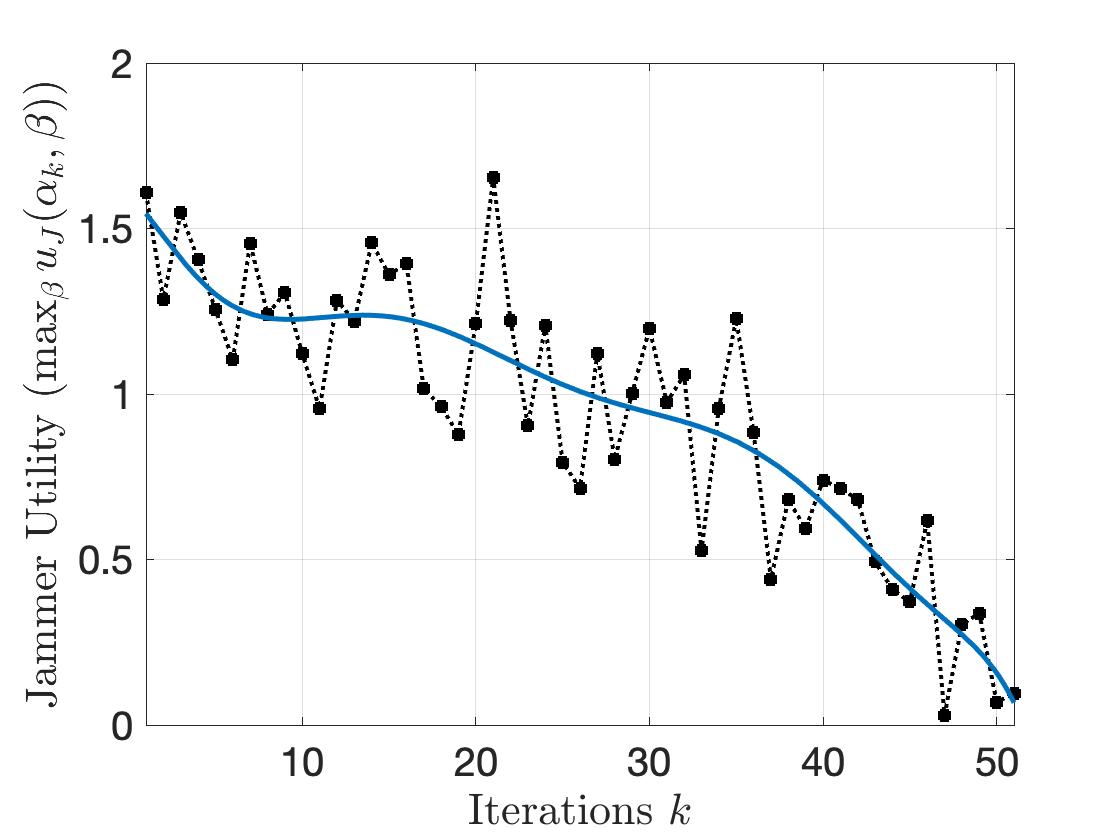}
   \caption{}
   \label{fig:jammer-utility-decay} 
\end{subfigure}
\caption{Adaptive ECCM mitigates jammer utility as well as obtains a more accurate estimate of the jammer's utility over time $\time$. In sub-figure (a), the radar's estimate of the jammer's coefficients improves with $k$. As seen in (b), the jammer utility decreases with $\time$ as the radar's estimate of $\param$ improves. }
\label{fig:plots}
\end{figure}
Theorem~\ref{thrm:rp} is our key IRL result that the radar uses to estimate the parameter $\param$ in the jammer's utility~\eqref{eqn:jammer_utility_rep}. The inequalities \eqref{eqn:AFT-ineq} are called Afriat's inequalities in literature. The feasibility of the inequalities~\eqref{eqn:AFT-ineq} is both necessary and sufficient for the dataset $\dataset_\time$ to be generated by a utility maximizer. The feasible set $\paramset_\time$~\eqref{eqn:set-of-feasible-param} comprises the set of feasible parameters that rationalize the radar's dataset $\dataset_\time$~\eqref{eqn:dataset-def}. 

In the radar context, the radar uses Afriat's theorem to {\em identify} if the jammer is cognition. Theorem~\ref{thrm:rp} assumes the radar knows the parameter $\lambda$ in \eqref{eqn:jammer_utility_rep}. If the feasibility inequalities~\eqref{eqn:AFT-ineq} have a viable solution,  Theorem~\ref{thrm:rp} provides a set-valued reconstruction of the jammer's utility parameter $\param$ via \eqref{eqn:set-of-feasible-param}. However, the radar's optimization~\eqref{eq:pap-irl-general} requires a point-valued estimate of the jammer's utility and not a set-valued estimate. Hence, for the adaptive ECCM to be well-posed, we assume the radar uses the {\em max-margin} estimate of the jammer's utility parameter $\param$:
\begin{align}
&\paramestiter  = \updateparam(\dataset_\time) =  \max_{\param'\in\paramset_\time}\margin(\param')\label{eqn:IRL-update}\\
&\margin(\param')  =
\min_{s,t\in\{1,\ldots,\time\}} \utility_{\param'}(\response_s)-\utility_{\param'}(\response_t)+\lambda \probe_t' (\response_s-\response_t),\label{eqn:margin}
\end{align}
where $\paramset_\time$ is computed via the feasibility test~\eqref{eqn:set-of-feasible-param}. The variable $\margin(\cdot)$ is the margin of the IRL feasibility test~\eqref{eqn:abstract_IRL_utility} given dataset $\dataset_\time$. Intuitively, the margin indicates how far is a candidate parameter estimate $\param'$ from failing the IRL test for utility maximization behavior; see \cite{PKB22-jrnl} for a elaborate discussion on the margin of set-valued IRL estimators. Our choice of the max-margin estimate for the jammer's utility is corroborated by similar approaches in~\cite{SVM} and also IRL~\cite{RAT06} for MDPs.

\section{Numerical Experiment. Adaptive ECCM}
\label{sec:numerical}
We now illustrate our adaptive ECCM scheme in Protocol~\ref{prtcl:adaptive-ECCM} via a simulation-based numerical experiment. Our numerical experiment uses the following parameters:
\begin{align}
&\text{Jammer action: }  \response_k\in\reals^{4}_{+},~\|\response_{\time}\|_2\leq 1\nonumber\\
&\text{Jammer utility: } \jammerUtility(\probe,\response)=-\response'\operatorname{diag}(\theta)\beta+\lambda\,\probe'\response,\label{eqn:sim-jammer-utility}\\
&\text{Radar action: }\probe_k\in\reals^{4}_{+},~\|\probe_{\time}\|_2\leq 1,\nonumber\\
&\text{Radar utility: } \radarUtility(\probe,\response)=
    \frac{\probe'\operatorname{diag}\response\probe}{\probe'\operatorname{diag}(\response)\probe+\response'\response} - \delta~~\jammerUtilityest(\probe,\response),\label{eqn:sim-radar-utility}\\
    &\text{Jammer utility}  \text{ parameter}:  \nonumber\\
    &\param=[\param_1~\param_2~\param_3~\param_4],~\param_i \sim \operatorname{Unif}([0,1]),~i=1,2,3,4,~\lambda=0.5\nonumber\\
    &\text{Radar's initial guess for $\param$:}\nonumber\\
    & \hat\param_{0}=[0.5,\,0.5,\,0.5,\,0.5]\nonumber
\end{align} 
Figure \ref{fig:plots} shows the performance of the radar's adaptive ECCM scheme of Protocol~\ref{prtcl:adaptive-ECCM} over $\horizon = 50$ interactions between the radar and jammer. Figure \ref{fig:plots} plots the IRL estimation error $\|\param-\hat{\param}_{\time}\|_{2}$ and jammer utility $\jammerUtility(\probe_k^{*},\response_k^{*})$ over time. We observe that the jammer's utility decreases with $\time$, and the distance between the radar's IRL estimate of the jammer's utility parameter $\paramestiter$ and true parameter $\param$ decreases with time. This trend is expected since (1) the size of the set-valued IRL parameter estimate $\paramset_\time$~\eqref{eqn:set-of-feasible-param} reduces with number of radar-jammer interactions and (2) the radar trades-off between maximizing its own utility and minimizing the jammer's utility~\eqref{eqn:sim-radar-utility}.

% \begin{figure}[]
%     \centering
%     \includegraphics[width=0.8\linewidth]{ICASSP 2023 - PAP-IRL-Hybrid/ErrorIterations.jpg}
%     \caption{Estimation error decreases as the iterations increases. We used a $6^{\text{th}}$ order polynomial to fit the decrease}
%     \label{fig:CoeffError}
% \end{figure}
% \begin{figure}[!ht]
%     \centering
%     \includegraphics[width=\linewidth]{CombinedPlot.jpg}
%     \caption{}
%     \label{fig:JammerUtilityDecay}
% \end{figure}

\section{Conclusion}
This paper provides a principled approach to adaptive ECCM where the radar needs to learn the jammer's utility over time. {\em How to learn a jammer's ECM tactics and mitigate ECM via ECCM at the same time?} Our key adaptive ECCM result is Protocol~\ref{prtcl:adaptive-ECCM}. The main idea is to choose a myopic action that optimizes radar performance based on current estimate of the jammer's utility followed by updating the estimate of the jammer's utility. To do so, we use the concepts of principal agent problem (PAP) from contract theory and revealed preference-based inverse reinforcement learning tools from micro-economics. In terms of future extensions, it is of interest to investigate the convergence properties of the adaptive ECCM scheme based on hybrid PAP-IRL. We shall also consider the case where the radar observes a noisy response from the jammer, and propose an adaptive ECCM algorithm where the IRL block is replaced with an inverse detector.  
% We jointly maximized the radar utility and learned the jammer's utility by optimizing the dual objective problem of PAP and IRL. By doing this we reduced the information asymmetry between the radar and the jammer. By Afriat's Theorem we showed the existence of the feasible set of parameters for jammer's utility. 
%\vfill
%\pagebreak

%\section{COPYRIGHT FORMS}
% \label{sec:copyright}
% \cite{HA87-CONTRACT}
% You must submit your fully completed, signed IEEE electronic copyright release
% form when you submit your paper. We {\bf must} have this form before your paper
% can be published in the proceedings.

\clearpage
%\vfill\pagebreak

\label{sec:refs}

\bibliographystyle{unsrt_abbrv_custom}
\bibliography{PAP_IRL}

\end{document}